\documentclass[sigconf]{acmart}

\usepackage{bm}
\usepackage{multirow}
\usepackage{subfigure}
\usepackage{algorithmic}
\usepackage[ruled, linesnumbered]{algorithm2e}
\usepackage{booktabs} 
\usepackage{amsmath}
\usepackage{amssymb}
\usepackage{amsmath}
\usepackage{xcolor}
\usepackage{multirow}
\usepackage{graphicx}
\usepackage{subfigure}
\usepackage{amsfonts}
\usepackage{balance}

\newcommand{\fullname}{\emph{Adaptive Graph Convolutional Network~(AGCN)}}
\newcommand{\shortname}{\emph{AGCN}}

\AtBeginDocument{%
  \providecommand\BibTeX{{%
    \normalfont B\kern-0.5em{\scshape i\kern-0.25em b}\kern-0.8em\TeX}}}

\copyrightyear{2020}
\acmYear{2020}
\setcopyright{acmlicensed}\acmConference[SIGIR '20]{Proceedings of the 43rd International ACM SIGIR Conference on Research and Development in Information Retrieval}{July 25--30, 2020}{Virtual Event, China}
\acmBooktitle{Proceedings of the 43rd International ACM SIGIR Conference on Research and Development in Information Retrieval (SIGIR '20), July 25--30, 2020, Virtual Event, China}
\acmPrice{15.00}
\acmDOI{10.1145/3397271.3401144}
\acmISBN{978-1-4503-8016-4/20/07}

\begin{document}
\fancyhead{}

\title{
Joint Item Recommendation and Attribute Inference: \\An Adaptive Graph Convolutional Network Approach
}

\author{Le Wu}
\affiliation{Key Laboratory of Knowledge Engineering with Big Data, Hefei University of Technology}
\affiliation{School of Computer Science and Information Engineering, Hefei University of Technology}
\email{lewu.ustc@gmail.com}

\author{Yonghui Yang}
\affiliation{Key Laboratory of Knowledge Engineering with Big Data, Hefei University of Technology}
\affiliation{School of Computer Science and Information Engineering, Hefei University of Technology}
\email{yyh.hfut@gmail.com}

\author{Kun Zhang}
\affiliation{Key Laboratory of Knowledge Engineering with Big Data, Hefei University of Technology}
\affiliation{School of Computer Science and Information Engineering, Hefei University of Technology}
\email{zhang1028kun@gmail.com}

\author{Richang Hong}
\affiliation{Key Laboratory of Knowledge Engineering with Big Data, Hefei University of Technology}
\affiliation{School of Computer Science and Information Engineering, Hefei University of Technology}
\email{hongrc.hfut@gmail.com}

\author{Yanjie Fu}
\affiliation{College of Engineering and Computer Science, University of Central Florida}
\email{yanjie.fu@ucf.edu}
\author{Meng Wang}
\authornote{Corresponding Author}
\affiliation{Key Laboratory of Knowledge Engineering with Big Data, Hefei University of Technology}
\affiliation{School of Computer Science and Information Engineering, Hefei University of Technology}
\email{eric.mengwang@gmail.com}

\begin{abstract}
In many recommender systems, users and items are associated with attributes, and users show preferences to items.
The attribute information describes users'~(items') characteristics and has a wide range of applications, such as user profiling, item annotation, and feature-enhanced recommendation. As annotating user~(item) attributes is a labor intensive task, the attribute values are often incomplete with many missing attribute values. Therefore, \emph{item recommendation} and \emph{attribute inference} have become two main tasks in these platforms. Researchers have long converged that user~(item) attributes and the preference behavior are highly correlated. Some researchers proposed to leverage one kind of data for the remaining task, and showed to improve performance. Nevertheless, these models either neglected the incompleteness of user~(item) attributes or regarded the correlation of the two tasks with simple models, leading to suboptimal performance of these two tasks.

To this end, in this paper, we define these two tasks in an attributed user-item bipartite graph, and propose an \fullname~approach for joint item recommendation and attribute inference. The key idea of \shortname~is to iteratively perform two parts: 1) Learning graph embedding parameters with previously learned approximated attribute values to facilitate two tasks; 2) Sending the approximated updated attribute values back to the attributed graph for better graph embedding learning. Therefore, \shortname~could adaptively adjust the graph embedding learning parameters by incorporating both the given attributes and the estimated attribute values, in order to provide weakly supervised information to refine the two tasks. Extensive experimental results on three real-world datasets clearly show the effectiveness of the proposed model.

\end{abstract}



\maketitle

\section{Introduction}
Collaborative Filtering~(CF) is one of the most popular approaches for recommender systems, which suggests personalized \emph{item recommendation} by collaboratively learning user and item embeddings  from user-item behavior~\cite{mnih2008PMF,UAI2009BPR}. However, as users' behavior data are usually sparse, CF suffers from the cold-start problem. A possible solution to solve the cold-start problem is to introduce auxiliary data for recommendation, such as text~\cite{kodakateri2009conceptual,poirier2010text-based,sun2020dual}, social networks~\cite{tkde2017modeling,SIGIR2019DiffNet}, user~(item) features and so on.  Among them,  user~(item) features are very common in most social platforms. E.g., users have pages to show their personal profiles, including  age, gender, occupation and so on. Items are annotated with tag information, such as movies are annotated with multiple genres~(action, adventure, romance and so on).  As user and item attributes describe user and item content information, given the complete attribute feature vector of each user~(item), attribute enhanced collaborative filtering models have been proposed to tackle the cold-start problem~\cite{jmlr2009SVD,TIST2012LIBFM}. These models extended CF with additional bias terms or modeled feature interactions for preference prediction.

For most attribute enhanced recommendation models, a simple assumption is that the attribute values are complete. However, as annotating user~(item) attributes is a labor intensive task,  many attribute values are often incomplete with missing attribute values. E.g., some users are unwilling to fill their gender attributes while others do not show their ages. Therefore, \emph{attribute inference} has become a main task in these recommendation platforms, with applications such as user~(item) search with a particular attribute, user profiling, item semantic understanding, and so on. By treating user-item interaction behavior as an attributed graph with partially known attributes, semi-supervised learning models infer each missing attribute of a user~(item) by collectively modeling node attributes and graph structure of different nodes~\cite{zhu2002LP,belkin2006manifold,zhu2003semi,ICLR2017Semigcn}.  Most classical semi-supervised approaches modeled label correlations in the graph, such as label propagation~\cite{zhu2002LP} or graph regularization~\cite{belkin2006manifold}.
Recently, Graph Convolutional Networks~(GCN) have shown huge success for semi-supervised learning, which iteratively combines node attributes and graph structures with convolutional operations for learning a dense node embedding vector.
Then, the learned user~(item) embeddings are directly fed into a simple linear model for attribute inference~\cite{ICLR2017Semigcn,NIPS2017inductive,AAAI2018deeper}.

With the user-item behavior data and incomplete user~(item) attributes, most researchers focused on leveraging one kind of data for the remaining task, i.e., attribute enhanced item recommendation or semi-supervised attribute inference with the user-item bipartite graph.  We argue that these two tasks are correlated and should not be modeled in an isolated way. On one hand, as the attribute values are incomplete,  most attribute enhanced recommendation algorithms take the inferred attribute values as input  for item recommendation. On the other hand, users' behaviors to items could well reflect user and item attributes~\cite{pnas2013private}. As both the behavior data and the attribute data are sparse, modeling these two tasks together would allow them to mutually reinforce each other. Some works made preliminary attempts to jointly model the correlation of these two tasks~\cite{gong2014joint,WWW2017BLA}. By reformulating the available data as an attributed graph, these models usually relied on classical graph based models to predict these two tasks, such as label propagation and graph link prediction, and are optimized in a joint loss function that combines two tasks. These models showed superior performance compared to modeling these two tasks separately. However, these models relied on classical shallow semi-supervised learning models,  and the performance is still unsatisfactory.

In this paper, we tackle the joint item recommendation and attribute inference under GCN based approaches. We choose GCN as the base model as it naturally inherits the power of deep learning models for automatically representation learning ability, and shows superior performance for various graph-based tasks~\cite{ICLR2017Semigcn}, such as item recommendation in the user-item bipartite graph~\cite{ICLR2017GCMC,SIGIR2019NGCF,AAAI2020revisiting}, node classification and clustering~\cite{ICLR2017Semigcn,abu2018ngcn}. Given the user-item attributed graph with some missing data, a naive idea for jointly modeling these two tasks is to apply  GCNs to learn the user and item embeddings, with the final optimization function consists of combining attribute inference loss and item recommendation loss. However, most GCNs assume the input feature is complete and could not tackle the missing attribute values. Therefore, the key challenge lies in the incompleteness of the attributes, which appears in both the input data and treated as the prediction variable. In other words, the learning process of GCNs relies on the complete input attribute values to inform user~(item) embedding, while the learned user~(item) embedding are beneficial for the  missing attribute inference. Simply filling the missing attributes with  static precomputed values would lead to weak local optimal performance as the input data is noisy.

To tackle this challenge, we propose an Adaptive Graph Convolutional Network~(AGCN) for both item recommendation and attribute inference. The main idea of AGCN is that, instead of sending fixed values to complete the missing attributes into GCN, AGCN adaptively  performs the two steps: adjusting the graph embedding learning parameters with estimated attribute values to facilitate both attribute inference and item recommendation, and updating the graph input with the newly approximated updated attribute values. Therefore, AGCN could adaptively learn the best GCN parameters to sever both tasks.  Finally, we conduct extensive experimental results on three real-world datasets. The experimental results clearly show the effectiveness of our proposed model for both item recommendation and attribute inference. E.g., our proposed model shows about 7\% improvement for item recommendation and more than 10\% improvement for attribute inference compared to the best baselines.

\section{Related Work}
CF has been widely used in most recommender systems due to its relatively high performance with easy to collect data~\cite{hu2008collaborative,marlin2009collaborative,tist2016relevance}. Classical latent factor based models relied on matrix factorization for user and item embedding learning~\cite{mnih2008PMF,koren2009MF,cheng2018aspect,cheng20183ncf}. As most users implicitly express their item preferences, Bayesian Personalized Ranking~(BPR) was proposed with a ranking based loss function to deal with the implicit feedback~\cite{UAI2009BPR}. In practice, CF based models suffer from the cold-start problem and could not perform well when users have limited rating records~\cite{park2009pairwise}. To tackle the data sparsity issue, many efforts have been devoted to incorporate auxiliary information in CF based models, such as user~(item) attributes~\cite{ICDM2010FM}, item content~\cite{AAAI2016VBPR}, social network~\cite{SIGIR2018attentive,sigir2019diffnet}, and so on. Among them, attribute enhanced CF are widely studied as the attribute information are easy to collect in most platforms. Researchers proposed to mimic the latent factor distribution with the associated features of users and items~\cite{agarwal2009regression}. SVDFeature extended over classical latent factor based models with additional bias terms, which are learned from the associated attributes~\cite{jmlr2009SVD}. Factorization machines modeled pairwise interactions between all features and was a generalized model since they can mimic most factorization models with feature engineering~\cite{ICDM2010FM,TIST2012LIBFM}. All these feature enhanced CF models assume that the attribute information is complete. However, in the real-world, user and item attributes are incomplete with many missing values. As these models could not tackle the missing feature value issue, a preprocessing step is usually adopted to fill the missing values, such as each missing attribute value is filled by the average value, or a computational model to predict the missing values at first~\cite{ICML2014probabilistic,cikm2019adaptive}. Instead of using preprocessing step to tackle the missing attribute problem, we design a model that learns attribute inference and item recommendation at the same time.

Recently, GCNs have shown huge success for graph representation learning and related applications~\cite{ICLR2017Semigcn,ICLR2017GCMC,NIPS2017inductive}. As the user-item behavior could be naturally regarded as a graph structure, researchers proposed graph based recommendation models for better user and item embedding learning ~\cite{KDD2018PinSage,ICLR2017GCMC,SIGIR2019NGCF,SIGIR2019DiffNet}. E.g., PinSage is a state-of-the-art content based GCN model for item recommendation, which incorporates both graph structure as well as node features for representation learning~\cite{KDD2018PinSage}. Given the user and item free embedding as input, NGCF performs neural graph collaborative filtering, with each user's~(item's) node embedding is recursively updated from the local neighbors' embeddings~\cite{SIGIR2019NGCF}. LR-GCCF is a general GCN based CF model with simplified linear graph convolution operations and residual learning between different layers, and shows better recommendation performance without annoying non-linear activation function tuning process. Most current GCN based recommendation models either fall in to the CF category or the content based recommendation. Our work differs from these models as we simultaneously consider the propagation of collaborative signals and node attributes with missing attributes.

Our work is also closely related to attribute inference on social platforms. We could cast each attribute inference as a supervised  classification~(regression) problem by treating the remaining attributes as input features, and attribute values of the current attribute as labels~\cite{agarwal2009regression}. However, this simple idea fails when the attribute information is incomplete with a large number of missing values. In fact, sociologists have long converged that users' behaviors and their attributes are highly correlated, termed as the homophily effect~\cite{mcpherson2001birds,pnas2013private}.
Instead of the two stage framework of first user~(item) embedding learning followed by the supervised attribute prediction, a more intuitive idea is to directly perform semi-supervised attribute inference given the user-item behavior graph. Label propagation investigated how to disseminate available attribute information in the graph for attribute inference~\cite{zhu2002LP}. However,
the model neglected the correlation of different attributes in the modeling process. Graph manifold was a general solution to semi-supervised learning and it extended over classical supervised learning models with a graph Laplacian regularization term~\cite{belkin2006manifold}. GCNs have been the state-of-the-art models for graph based tasks, such as node classification, node clustering, and community detection~\cite{ICLR2017Semigcn,chiang2019cluster,jin2019graph}. GCNs could encode the graph structure for representation learning by taking the node features and the graph structure, with a supervised target for nodes with labels~\cite{ICLR2017GCMC,NIPS2017inductive,AAAI2018deeper}. We borrow the advantage of GCNs for semi-supervised learning, and design an adaptive graph convolutional model to tackle the missing attribute problem.

Due to the reinforcement relationship between these two tasks, some works have considered to jointly predict these two tasks in a unified framework~\cite{gong2014joint,WWW2017BLA}.
These models have reformulated the available data as an attributed graph, and designed joint optimization functions from these two tasks based on the classical item recommendation model and attribute inference model. E.g, researchers summarized various classical link prediction and attribute inference models~\cite{gong2014joint}. BLA iteratively performed label propagation for attribute inference, and biased random walks with learned attributes for item recommendation~\cite{WWW2017BLA}. Despite the improved performance compared to single task modeling, these models relied on the classical shallow graph based models or random walk based approaches for the two prediction tasks. Therefore, these models suffer from huge time complexity and could not well encode the complex global structure to facilitate these two tasks.

\section{Problem Definition}
In a recommender system, there are two sets of entities: a userset~{\small$U$~($|U|\!=\!M$)}, and an itemset~{\small$V$~($|V|\!=\!N$)}.  Since implicit feedback is very common, we use a rating matrix {\small $\mathbf{R}\in\mathbb{R}^{M\times N}$} to denote users' implicit feedback to items,  with $r_{ai}\!=\!1$ if user $a$ has connection with item $i$, otherwise it equals 0. Besides the rating information, users and items are often associated with attributes. We use {\small $\mathbf{X}\in\mathbb{R}^{d_x\times M}$}  and {\small $\mathbf{Y}\in\mathbb{R}^{d_y\times N}$} to denote user and item attribute matrix,  where $d_x$ and $d_y$ are the dimension of user and item attributes, respectively.

In the real world, user and item attributes are often incomplete. We use {\small $\mathbf{A}^X\in\mathbb{R}^{d_x\times M}$} as the user attribute indication matrix, with $a^X_{iu}=1$ denotes the $i^{th}$ attribute of user $u$ is available. Under this circumstance, $x_{iu}$ shows the detailed $i^{th}$ value of user $u$. In contrast, $a^X_{iu}=0$ denotes the $i^{th}$ attribute of user $u$ is missing. Similarly, we use {\small $\mathbf{A}^Y\in\mathbb{R}^{d_y\times N}$} to denote the item attribute indication matrix, with $a^Y_{jv}=0$ representing the $j^{th}$ attribute of item $v$ is available, otherwise it equals 0. Then, the problem we study in this paper is defined as:

\begin{definition}
    In a recommender system, given the user set $U$, item set $V$, user-item preference matrix {\small$\mathbf{R}\in\mathbb{R}^{M\times N}$}, user attribute matrix {\small $\mathbf{X}\in\mathbb{R}^{d_x\times M}$} and attribute value indication matrix {\small $\mathbf{A}^X\in\mathbb{R}^{d_x\times M}$}, item attribute matrix {\small $\mathbf{Y}\in\mathbb{R}^{d_y\times N}$} along with item attribute value indication matrix {\small $\mathbf{A}^Y\in\mathbb{R}^{d_y\times N}$}, our goals are to recommend items to users and predict the missing attribute values of either users or items. These two goals can be formulated as follows:
    \begin{itemize}
      \item \emph{Item Recommendation}: This goal aims at predicting users' preferences to unrelated items as: $\hat{\mathbf{R}}=f(\mathcal{G}=[U,V,\mathbf{R},\mathbf{X},\mathbf{Y},\mathbf{A}^X,$ $\mathbf{A}^Y])$, where $\hat{\mathbf{R}}\in\mathbb{R}^{M\times N}$ denotes the predicted rating matrix.
      \item \emph{Attribute Inference}: This goal is to predict the missing attribute values of users or items as: $\hat{\mathbf{X}}=g^X(\mathcal{G})$, $\hat{\mathbf{Y}}=g^Y(\mathcal{G})$, where $\hat{\mathbf{X}}$ and $\hat{\mathbf{Y}}$ mean the predicted user attribute matrix and item attribute matrix, respectively.
\end{itemize}
\end{definition}

\section{The Proposed Model}

\begin{small}
\begin{figure*} [htb]
  \begin{center}
    \includegraphics[width=0.95\textwidth]{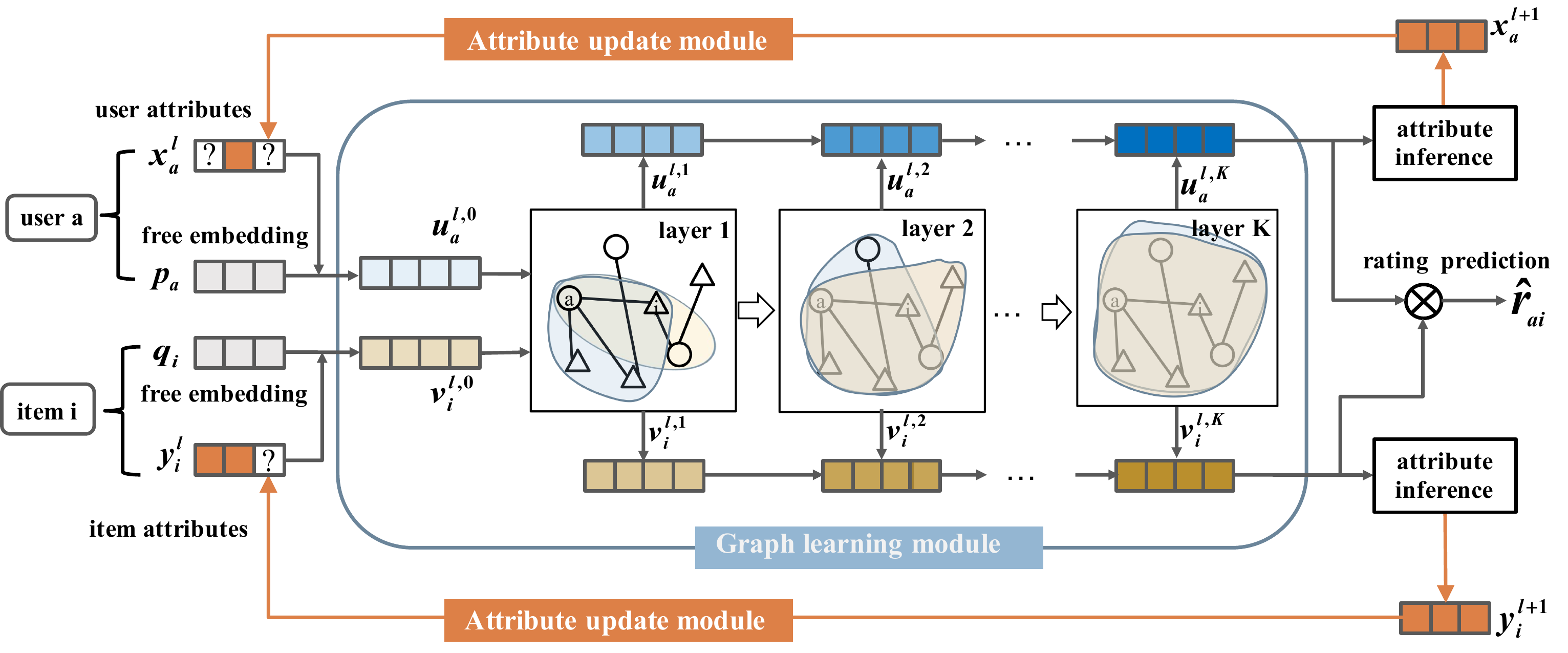}
  \end{center}
    \vspace{-0.5cm}
  \caption{The overall framework of our proposed model.}\label{fig:framework}
   \vspace{-0.3cm}
\end{figure*}
\end{small}

In this section, we would introduce our proposed \shortname~ model for joint item recommendation and attribute inference.  We would first introduce the overall architecture of the proposed model, followed by the model optimization process.

\subsection{Overall Architecture}
Figure~\ref{fig:framework} illustrates the overall architecture of our proposed \shortname, which contains two iterative main modules:
\textit{graph learning module} and \textit{attribute update module}. At each iteration $l$, the graph learning module takes the  predicted $l$-th attribute values as input, and learns the network parameters. By feeding the learned graph parameters and user~(item) embeddings into the attribute update module, this module is designed to infer missing user~(item) attribute values. After that, the missing attributed values are updated as the $(l+1)^{th}$ approximated
attribute value and are sent back to the graph learning module. These two modules are iterated from $l=0$ until
the model converges.

\subsection{Graph Learning Module}
At each iteration $l$, the graph learning module contains two components: the embedding fusion layer and the embedding propagation layers. The embedding fusion layer fuses each node's free embedding, as well as approximated attribute values at the $l^{th}$ iteration.  The embedding propagating layers propagate the
fused embeddings in order to capture the higher-order graph structure for user~(item) representation learning.

\textbf{Embedding Fusion Layer.} Similar to many embedding based models, we use {\small$\mathbf{P}\in\mathbb{R}^{d\times M}$} and
{\small$\mathbf{Q}\in\mathbb{R}^{d\times N}$} to denote the free embedding matrix of user and item respectively. The free embedding matrix could capture the collaborative latent representations of users and items~\cite{UAI2009BPR,WWW2017NCF,AAAI2020revisiting}. Therefore, each user $a$'s free embedding is denoted as the $a^{th}$ column of {\small$\mathbf{P}$}, i.e, {$\mathbf{p}_a$}. Similarly, each item $i$'s free embedding is denoted as {$\mathbf{q}_i$}, which is the $i^{th}$ column of {\small $\mathbf{Q}$}.

At the $l^{th}$ iteration, the approximated user and item attribute input are denoted as {\small$\mathbf{X^l}\in\mathbb{R}^{d_x\times M}$} and {\small$\mathbf{Y^l}\in\mathbb{R}^{d_y\times N}$}, which is learned from the $(l-1)^{th}$ attribute update module. Given the approximated user attribute vector $\mathbf{x}^l_a$, and the approximated item attribute vector $\mathbf{y}^l_i$, we fuse the free embedding and the attribute embedding to get the fused embeddings as:

\begin{small}
\vspace{-2pt}
\begin{flalign}\label{eq:fusion}
\mathbf{u}^{l,0}_a &= [\mathbf{p}_a,\mathbf{x}^l_a\times{\mathbf{W}_u}],\\
\mathbf{v}^{l,0}_i &= [\mathbf{q}_i,\mathbf{y}^l_i\times{\mathbf{W}_v}],
\end{flalign}
\vspace{-2pt}
\end{small}

\noindent where {\small$\mathbf{W}_u\in\mathbb{R}^{d_x\times d_a}$} and
{\small$\mathbf{W}_v\in\mathbb{R}^{d_y\times d_a}$} are two transformation matrices that need to be learnt.
Along this line, \shortname~could represent users and items
with both the collaborative signal and the content signal.

Specifically, at the $0$-th iteration, we do not have any estimated missing attribute values learned in the attribute update module. In practice, we set ${\small\mathbf{X}^0}$ and ${\small\mathbf{Y}^0}$ as the average of the mean of the available attribute values:

\begin{small}
\begin{equation}\label{eq:attributes padding avg}
x^0_{fa} = \frac{\sum_{b=0}^{M-1}X_{fb}\times a^X_{fb}}{\sum_{b=0}^{M-1} a^X_{fb}}, \quad
y^0_{fi} = \frac{\sum_{j=0}^{N-1} y_{fj}\times a^Y_{fj}}{\sum_{j=0}^{N-1} a^Y_{fj}},
\end{equation}
\end{small}

\noindent where $a^x_{fb}$ is an element of user attribute indication matrix {\small$\mathbf{A}^X$}, and $a^Y_{fj}$
is an element of item indication matrix {\small$\mathbf{A}^Y$}.

\textbf{Embedding Propagation Layers.} In this part, we propagate users'~(items') fused embeddings to  capture the high-order proximity between users and items for better user and item embedding learning.
The inputs of this layer are the fusion user embeddings $\mathbf{u}^l_a$ and fusion item embeddings $\mathbf{v}^l_i$. Due to the embedding fusion, the user~(item) attributes will be propagated along with the user~(item) free embeddings in GCN.

To be specific, let $\mathbf{u}^{l,k}_a$ and $\mathbf{v}^{l,k}_i$ denote user $a$ and item $j$ embeddings in the $k^{th}$  layer. Their embeddings in the  $(k+1)^{th}$ layer can be defined by their fusion embeddings and the aggregation of corresponding connected items~(users) embeddings in $k^{th}$ layer.
This process can be formulated as follows:
\begin{small}
\vspace{-5pt}
\begin{flalign}
\mathbf{u}_a^{l,k+1} &= (\mathbf{u}_a^{l,k}+\sum_{j\in R_a} \frac{\mathbf{v}_j^{l,k}}{|R_a|})\times\mathbf{W}^{k+1}, \label{eq:gcn_uk}\\
\vspace{-0.2cm}
\mathbf{v}_i^{l,k+1} &= (\mathbf{v}_i^{l,k}+\sum_{b\in S_i} \frac{\mathbf{u}_b^{l,k}}{|S_i|})\times\mathbf{W}^{k+1},  \label{eq:gcn_vk}
\end{flalign}
\vspace{-5pt}
\end{small}

\noindent where {\small$R_a=\{i|R_{ai}\!=\!1\}\subseteq I$} is the item set that user $a$ interacts with.
Similarly, {\small$S_i=\{a| R_{ai}\!=\!1\}\subseteq U$} is the user set who interact with item $i$.  {\small$\mathbf{W}^{k+1}\in\mathbb{R}^{(d+d_a)\times (d+d_a)}$} is a transformation matrix in the $k^{th}$  layer propagation of user-item graph. Please note that, in the above equations, we use the linear convolutional operations without any non-linear activations, as this simple operation shows simplicity, and outperforms its counterparts with non-linear transformations in CF~\cite{AAAI2020revisiting}. Thus, each user's~(item's) k-order neighbors embedding will be propagated at $(k+1)^{th}$ embedding propagation layer.

To better illustrate the propagation process, we formulate the embedding propagation as a matrix form. Let {\small$\mathbf{A}$} denote the adjacency matrix of user-item bipartite graph with {\small$(M+N)$} nodes:
\begin{small}
\vspace{-2pt}
\begin{flalign}\label{eq:adj_matrix}
\mathbf{A}=\left[\begin{array}{cc}
\mathbf{R} & \mathbf{0^{N\times M}}\\
\mathbf{0^{M\times N}} & \mathbf{R}^T
\end{array}\right].
\end{flalign}
\vspace{-2pt}
\end{small}

Let $\mathbf{U}^{l,k}$ and $\mathbf{V}^{l,k}$ denote the fusion embedding matrices of users and items at the $k^{th}$ layer in the $l$-th iteration, the fusion embedding matrix at $(k+1)^{th}$ layer is defined as:
\begin{small}
\vspace{-2pt}
\begin{flalign}\label{eq:matrix_propagation}
\left[\begin{array}{c}
\mathbf{U}^{l,k+1} \\
\mathbf{V}^{l,k+1}\end{array}\right]
=(\left[\begin{array}{c}
\mathbf{U}^{l,k} \\
\mathbf{V}^{l,k}\end{array}\right]+
\mathbf{D}^{-0.5}\mathbf{A}\mathbf{D}^{0.5}\times
\left[\begin{array}{c}
\mathbf{U}^{l,k} \\
\mathbf{V}^{l,k}\end{array}\right])\times \mathbf{W}^{k+1},
\end{flalign}
\vspace{-2pt}
\end{small}

\noindent where $\mathbf{D}$ is the degree matrix of $\mathbf{A}$, which aims to smooth the clustered neighbors' embeddings. This matrix form of propagation takes all users~(items) into propagation and updates the corresponding fusion matrices simultaneously in an efficient way.

\subsection{Attribute Update Module}
This module is consist of two parts: prediction part and attribute update part.  The prediction part predicts the
user preference for item recommendation and attribute inference. The attribute update part updates the missing user and item attributes according to the  inferred attribute values.

\textbf{Prediction Part.}
Given the propagation depth $K$, we could obtain the final embedding $\mathbf{u}^{l,K}_a$ for user $a$ and final embedding $\mathbf{v}^{l,K}_i$ for item $i$ after $K$ iterative propagation layers.
Thus, the preference of user $a$ for item $i$ can be defined as follows:

\begin{flalign}\label{eq:rating prediction}
\mathbf{\hat{r}_{ai}} = <\mathbf{u}^{l,K}_a,\mathbf{v}^{l,K}_i>,
\end{flalign}

\noindent where <,> denotes vector inner product operation.

For attribute inference task, we also make full use of the learned user and item embeddings output by propagation layers.  We leverage the final embedding $\mathbf{u}^{l,K}_a$ for user $a$ and final embedding $\mathbf{v}^{l,K}_i$ for item $i$ to infer the missing values of attributes as follows:
\begin{small}
\begin{flalign} \label{eq:inference}
\begin{aligned}
\mathbf{\hat{x}}_a &= softmax(\mathbf{u}^{l,K}_a\times \mathbf{W}_x ), \\
\mathbf{\hat{y}}_i &= softmax(\mathbf{v}^{l,K}_i\times \mathbf{W}_y ),
\end{aligned}
\end{flalign}
\end{small}

\noindent where {\small$\mathbf{W}_x\in\mathbb{R}^{(d+d_a)\times d_x}$} and {\small$\mathbf{W}_y\in\mathbb{R}^{(d+d_a)\times d_y}$} are the transformation matrices that need to be learnt.

\textbf{Attribute Update Part.} After inferring user attribute matrix $\hat{\mathbf{X}}$ and item attribute matrix $\hat{\mathbf{Y}}$, we adaptively update the missing attribute values with our inferred results. And the available attribute values would keep the same. Then, the $(l+1)^{th}$ update for user and item attributes can be formulated as follows:
\begin{small}
\vspace{-2pt}
\begin{flalign}\label{eq:update}
\begin{aligned}
\mathbf{X}^{l+1} &= \mathbf{X}^l\cdot \mathbf{A}^X+\hat{\mathbf{X}}\cdot{(\mathbf{I}^X-\mathbf{A}^X)},\\
\mathbf{Y}^{l+1} &= \mathbf{Y}^l\cdot \mathbf{A}^Y+\hat{\mathbf{Y}}\cdot{(\mathbf{I}^Y-\mathbf{A}^Y)}
\end{aligned}
\end{flalign}
\vspace{-2pt}
\end{small}

\noindent where {\small$\mathbf{I}^X \in\mathbb{R}^{d_x\times M}$} and {\small$\mathbf{I}^Y \in\mathbb{R}^{d_y\times N}$} are matrices  with all elements of 1. The updated user attribute matrix and item attribute matrix are sent back to the graph learning module for next training iteration~($(l+1)^{th}$ iteration) until convergence.

\subsection{Model Optimization}

Since the task in this paper contains two targets, the optimization also consists of two parts: item recommendation loss and attribute inference loss. We compute each part's objective function and combine them for final optimization.

\textbf{Item Recommendation Loss.}
We employ the pairwise ranking based BPR loss ~\cite{UAI2009BPR,koren2009MF}, which assumes that the observed items' prediction values should be higher than those unobserved. The objective function can be formulated as follows:
	\begin{equation}\label{eq:loss_rating}
		\arg\min_{\Theta_r}\mathcal{L}_r=\sum_{a=0}^{M-1}\sum\limits_{(i,j)\in D_a } -\mathbf{ln}\sigma(\hat{r}_{ai}-\hat{r}_{aj}) +\lambda||\mathbf{\Theta_1}||^2,
	\end{equation}

\noindent where $\sigma(x)$ is a sigmoid function, {\small $\Theta_r\!=\![\Theta_1,\Theta_2]$} is the parameter set in item recommendation, with $\Theta_1=[\mathbf{P},\mathbf{Q}]$ is user and item free embedding matrices, $\Theta_2=[[\mathbf{W}^k]_{k=1}^K,\mathbf{W}_u,\mathbf{W}_v]$. $\lambda$ is a regularization parameter that restraints the complexity of user and item free latent embedding matrices.
{\small$D_a=\{(i,j)|i\in R_a\!\wedge\!j\not\in R_a\}$} denotes the pairwise training data for user $a$. {\small$R_a$} represents the item set that user $a$ has rated.

\textbf{Attribute Inference Loss.}
Since attribute inference task can be formulated as a classification task, we use cross entropy loss~\cite{mannor2005cross,krizhevsky2012imagenet} to measure the error of inferred attributes as follows:

\begin{small}
	\begin{flalign}\label{eq:loss_attribute}
	\nonumber \arg\min_{\Theta_a}\mathcal{L}_a &= loss(\mathbf{X},\hat{\mathbf{X}},\mathbf{A}^X) + loss(\mathbf{Y},\hat{\mathbf{Y}},\mathbf{A}^Y)\\
	&= \sum\limits_{j=0}^{M-1} \sum_{i=0}^{d_x-1} -x_{ij}log{\hat{x}_{ij}}a^X_{ij} + \sum\limits_{j=0}^{N-1} \sum_{i=0}^{d_y-1} -y_{ij}log{\hat{y}_{ij}}a^Y_{ij},
	\end{flalign}
\end{small}

\noindent where {\small $\Theta_a=[\mathbf{W}_x,\mathbf{W}_y]$} is the parameter set in the attribute update module.

After obtaining the two tasks' objective functions, we set a parameter $\gamma$ to balance the item recommendation loss and attribute inference loss. The final optimization can be formulated as follows:
\begin{small}
	\begin{flalign}\label{eq:loss_all}
	\arg\min_{\Theta}
	\mathcal{L}&=\mathcal{L}_r + \gamma\mathcal{L}_a,
	\end{flalign}
\end{small}

\noindent where $\Theta=[\Theta_r,\Theta_a]$ are all parameters in the final objective function.
We implement the proposed model with TensorFlow\footnote{https://www.tensorflow.org}.
We leverage Adam optimizer to train the model.  As we emphasized in update attributes part, the training process is a recyclable process instead of end to end form. Since our inferred users~(items) attributes will update the initialized users~(items) attributes, we repeat this process until model convergence. The detailed algorithm for training is illustrated in Algorithm~\ref{alg: agcn}.

\begin{algorithm}
\renewcommand{\algorithmicrequire}{\textbf{Input:}}
\renewcommand\algorithmicensure {\textbf{Output:}}
\caption{\small{The Algorithm of \shortname}}\label{alg: agcn}
\begin{algorithmic}[1]
\REQUIRE User-item bipartite Graph $\mathcal{G}$; graph propagation depth $K$; ~~\\
\ENSURE Parameter $\Theta_r$ in graph learning module, $\Theta_a$ in attribute update module; ~~\\

\STATE Random initialize model parameters; \\
\STATE l=0;\\
\STATE Calculate initial user attribute {\small$\mathbf{X}^l$} and  {\small$\mathbf{Y}^l$}~(Eq.\eqref{eq:attributes padding avg});

\WHILE{not converged}
    \STATE Sample a batch of training data; \\
    \STATE Update {\small $\mathbf{u}^l$} and {\small $\mathbf{v}^l$} ~(Eq.\eqref{eq:fusion})
    \FOR {$k=0$ to $K-1$}
          \STATE Update $\mathbf{U}^{l,k+1}$ and $\mathbf{V}^{l,k+1}$~(Eq.\eqref{eq:matrix_propagation});\\
    \ENDFOR

    \STATE Predict rating preference~(Eq.\eqref{eq:rating prediction})
    \STATE Predict attribute values~(Eq.\eqref{eq:inference})
    \STATE Parameters update according to Eq. \eqref{eq:loss_all};
    \STATE Update the approximated attributes $\mathbf{X}^{l+1}$ and $\mathbf{Y}^{l+1}$~(Eq. \eqref{eq:update});
    \STATE l=l+1;
\ENDWHILE
\STATE Return $\Theta$.
\end{algorithmic}
\end{algorithm}

\begin{table*}[htb]
    \centering
	\setlength{\belowcaptionskip}{5pt} %
	\caption{The statistics of the three datasets ~(``s" means single-label attribute and ``m'' means multi-label attribute).}\label{tab:statistics}
	\vspace{-8pt}{
	\scalebox{0.88}{
    \begin{tabular}{c|c|c|c|c|c}
    \hline
    Dataset & Users & Items & Ratings & Sparsity & Attributes \\ \hline
    \multirow{3}{*}{Amazon-Video Games} & \multirow{3}{*}{31,207} & \multirow{3}{*}{33,899} & \multirow{3}{*}{300,003} & \multicolumn{1}{c|}{\multirow{3}{*}{99.972\%}} & \textit{Price(s)}: 0 $\sim$ 9 \\ \cline{6-6}
     &  &  &  & \multicolumn{1}{c|}{} & \textit{Platform(m)}: PC, Android, Mac, Linux, Xbox, PS4, Card, Board Games \\ \cline{6-6}
     &  &  &  & \multicolumn{1}{c|}{} & \textit{Theme(m)}: Advance, Shooting, Home Design, Music, Waltzes, Language, Puzzle \\ \hline
    \multirow{3}{*}{Movielens-1M} & \multirow{3}{*}{6,040} & \multirow{3}{*}{3,952} & \multirow{3}{*}{226,310} & \multirow{3}{*}{99.052\%} & \textit{Gender(s)}: man, woman \\ \cline{6-6}
     &  &  &  &  & \textit{Age(s)}: under 18, 18$\sim$24, 25$\sim$34, 35$\sim$44, 45$\sim$49, 50$\sim$55, 56+ \\ \cline{6-6}
     &  &  &  &  & \textit{Occupation(s)}: educator, artist, student, writer, scientist, doctor, engineer \\ \hline
    Movielens-20M &138,493 & 27,278 & 20,000,263 & 99.471\% & \textit{Genres(m)}: Action, Comedy, Drama, Sci-Fi, Thriller, Adventure, Thriller, War \\ \hline
    \end{tabular}}}
\end{table*}

\section{EXPERIMENTS}

\subsection{Experimental Settings}
\subsubsection{Datasets Description.} To evaluate the effectiveness of our proposed model, we conduct experiments on three public datasets: Amazon-Video Games, Movielens-1M, Movielens-20M.

\noindent\textbf{1) Amazon-Video Games.} Amazon datasets contain users' implicit feedbacks and rich product attributes~(such as size, price, platform and category)~\cite{mcauley2015image,he2016ups}. As the original dataset of product attributes are noisy with many attributes have appeared rarely, we choose a typical Amazon-Video Games dataset, and select the attributes that appear in top 25\%. After that, we have three attributes: price, platform and theme. For simplicity, we only consider binary attributes in this paper. In data pre-processing step, we split the price attribute into 10 intervals, and the price of each video game falls into one of the ten intervals. Therefore, the price attribute inference is a single-label classification problem. For the platform and the theme attributes, each video game could be played in multiple platforms, such as ``Xbox'', ``PS4'' and ``PC'. The video game may belong to different themes like ``Adventure'', ``Fitness'' and so on. Thus, the platform and theme attribute prediction problems can be treated as a multi-label classification problem. During data pre-processing, we use one-hot encoding to encode the single label attribute, and use binarization encoding to encode the multi-label attributes~\cite{choi2013method}.

\noindent\textbf{2) Movielens-1M.} Movielens~\cite{harper2015movielens} is a classical recommendation dataset with user attributes. As the original dataset is much denser than most CF datasets, we only select the ratings that equal 5 as users' liked movies to increases the sparsity of user-item behavior matrix.
The dataset contains three user attributes: gender, age, and occupation. All of three attributes are single-label attributes. During data pre-processing, we encode all these single-label attributes with one-hot encoding.

\noindent\textbf{3) Movielens-20M.} This dataset contains 20 million rating records of movies that users watched, as well as item attributes. We treat items that are rated by users as positive feedbacks. Each item belongs to several genres, such as action, sci-fi, and so on. Therefore, the genre attribute is a  multi-label attribute, and we encode them with binarization encoding.

For all datasets, we filter out users that have less than 5 rating records.
After data pre-processing, we randomly split historical interactions into training, validation, and test parts with the ratio of 8:1:1. For each type of attribute, we randomly delete the attribute values at the rate of $\alpha$.  Since the original datasets contain all the attribute values, we empirically set $\alpha=0.9$ and randomly delete 90\% attribute values. Therefore, the 90\% deleted attribute values are used for test.  Table \ref{tab:statistics} summarizes the statistics of the three datasets after pre-processing.

\begin{table*}[htb]
	\begin{small}
		\centering
		\setlength{\belowcaptionskip}{2pt}
		\caption{HR@N comparisons for item recommendation with different top-N values.}\label{tab:hr@n_topk}
		\vspace{-8pt}
		\resizebox{\textwidth}{!}{
		\begin{tabular}{|c|c|c|c|c|c|c|c|c|c|c|c|c|c|c|c|}
			\hline
			\multirow{2}{*}{Models} &
			\multicolumn{5}{|c|}{Amazon-Video Games} &\multicolumn{5}{|c|}{Movielens-1M} &\multicolumn{5}{|c|}{Movielens-20M} \\
			\cline{2-16}
			&N=10&N=20&N=30&N=40&N=50&N=10&N=20&N=30&N=40&N=50&N=10&N=20&N=30&N=40&N=50\\
			\hline
			BPR&0.0599&0.0921&0.1163&0.1367&0.1527&0.2076&0.2801&0.3405&0.3874&0.4273&0.2537&0.3008&0.3461&0.3848&0.4188\\
			\hline
			FM&0.0623&0.0967&0.1208&0.1412&0.1602&0.2151&0.2871&0.3470&0.3956&0.4372&0.2587&0.3082&0.3539&0.3934&0.4271\\
			\hline
			BLA&0.0657&0.1036&0.1321&0.1553&0.1784&0.2129&0.2866&0.3463&0.3956&0.4380&0.2631&0.3148&0.3612&0.3997&0.4331\\
			\hline	
			NGCF&0.0775&0.1194&0.1529&0.1799&0.2010&0.2211&0.2953&0.3561&0.4071&0.4507&0.2811&0.3353&0.3833&0.4241&0.4586\\
			\hline
            PinNGCF&0.0797&0.1218&0.1537&0.1785&0.2002&0.2207&0.2935&0.3555&0.4037&0.4448&0.2847&0.3400&0.3900&0.4312&0.4656\\
			\hline
		\emph{\textbf{AGCN}}&\textbf{0.0861}&\textbf{0.1340}&\textbf{0.1683}&\textbf{0.1958}&\textbf{0.2182}&\textbf{0.2261}&\textbf{0.3004}&\textbf{0.3607}&\textbf{0.4125}&\textbf{0.4517}&\textbf{0.2992}&\textbf{0.3520}&\textbf{0.3996}&\textbf{0.4393}&\textbf{0.4732}\\
			\hline
		\end{tabular}}
	\end{small}
\end{table*}

\begin{table*}[htb]
	\begin{small}
		\centering
		\setlength{\belowcaptionskip}{2pt}
		\caption{NDCG@N comparison  for item recommendation with different top-N values.}\label{tab:ndcg@n_topk}
		\vspace{-8pt}
		\resizebox{\textwidth}{!}{
		\begin{tabular}{|c|c|c|c|c|c|c|c|c|c|c|c|c|c|c|c|}
			\hline
			\multirow{2}{*}{Models} &
			\multicolumn{5}{|c|}{Amazon-Video Games}& \multicolumn{5}{|c|}{Movielens-1M} &\multicolumn{5}{|c|}{Movielens-20M} \\
			\cline{2-16}
			&N=10&N=20&N=30&N=40&N=50&N=10&N=20&N=30&N=40&N=50&N=10&N=20&N=30&N=40&N=50\\
			\hline
			BPR&0.0318&0.0401&0.0455&0.0496&0.0526&0.1903&0.2166&0.2384&0.2548&0.2680&0.2476&0.2583&0.2718&0.2841&0.2951\\
			\hline
			FM&0.0333&0.0415&0.0475&0.0516&0.0552&0.1953&0.2211&0.2426&0.2594&0.2730&0.2498&0.2601&0.2760&0.2886&0.2995\\
			\hline
			BLA&0.0350&0.0448&0.0511&0.0557&0.0601&0.1952&0.2214&0.2430&0.2599&0.2738&0.2556&0.2682&0.2821&0.2943&0.3051\\
			\hline	
			NGCF&0.0414&0.0523&0.0596&0.0651&0.0690&0.2006&0.2275&0.2493&0.2668&0.2811&0.2692&0.2830&0.2976&0.3106&0.3218\\
			\hline
            PinNGCF&0.0431&0.0540&0.0610&0.0660&0.0701&0.2014&0.2274&0.2497&0.2664&0.2664&0.2801&0.2906&0.3059&0.3190&0.3302\\
            \hline
		\emph{\textbf{AGCN}}&\textbf{0.0461}&\textbf{0.0585}&\textbf{0.0661}&\textbf{0.0716}&\textbf{0.0758}&\textbf{0.2065}&\textbf{0.2327}&\textbf{0.2544}&\textbf{0.2722}&\textbf{0.2853}&\textbf{0.2903}&\textbf{0.3028}&\textbf{0.3170}&\textbf{0.3295}&\textbf{0.3404}\\
			\hline
		\end{tabular}}
	\end{small}
\end{table*}

\begin{figure*}
\vspace{-0.2cm}
  \begin{center}
  \vspace{-0.2cm}
      \subfigure[Amazon-Video Games]{\includegraphics[width=55mm]{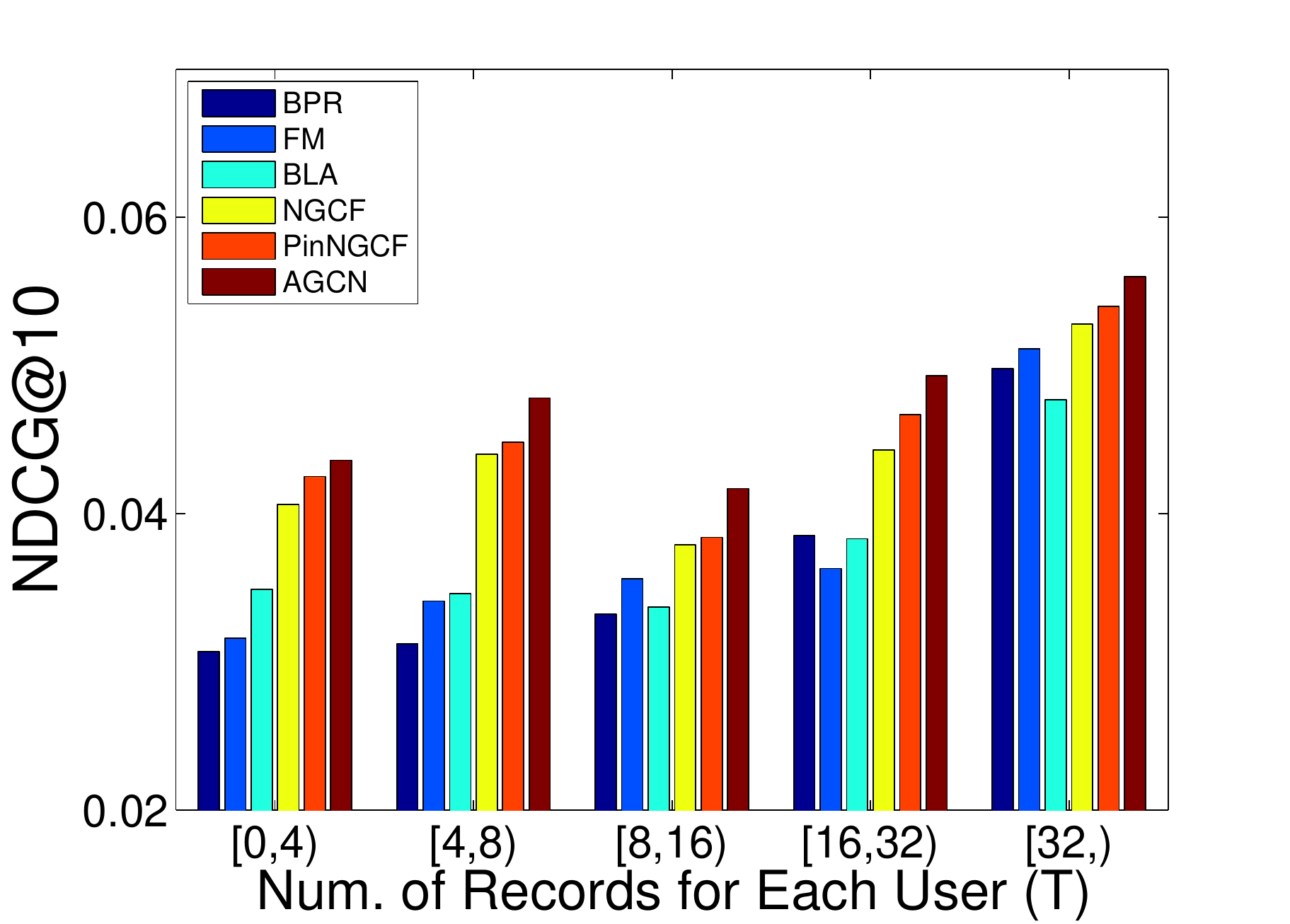}}
      \subfigure[Movielens-1M]{\includegraphics[width=55mm]{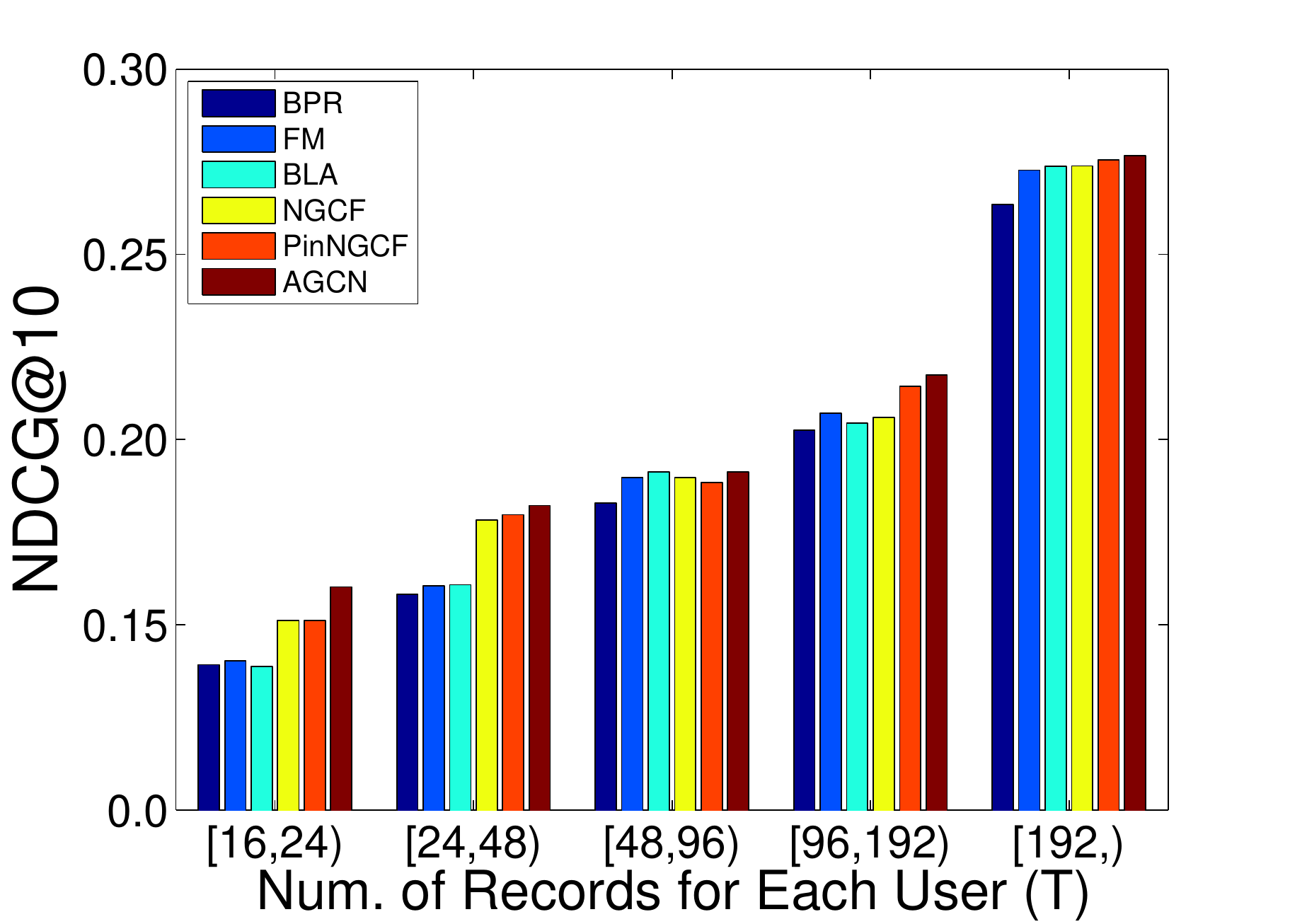}}
      \subfigure[Movielens-20M]{\includegraphics[width=55mm]{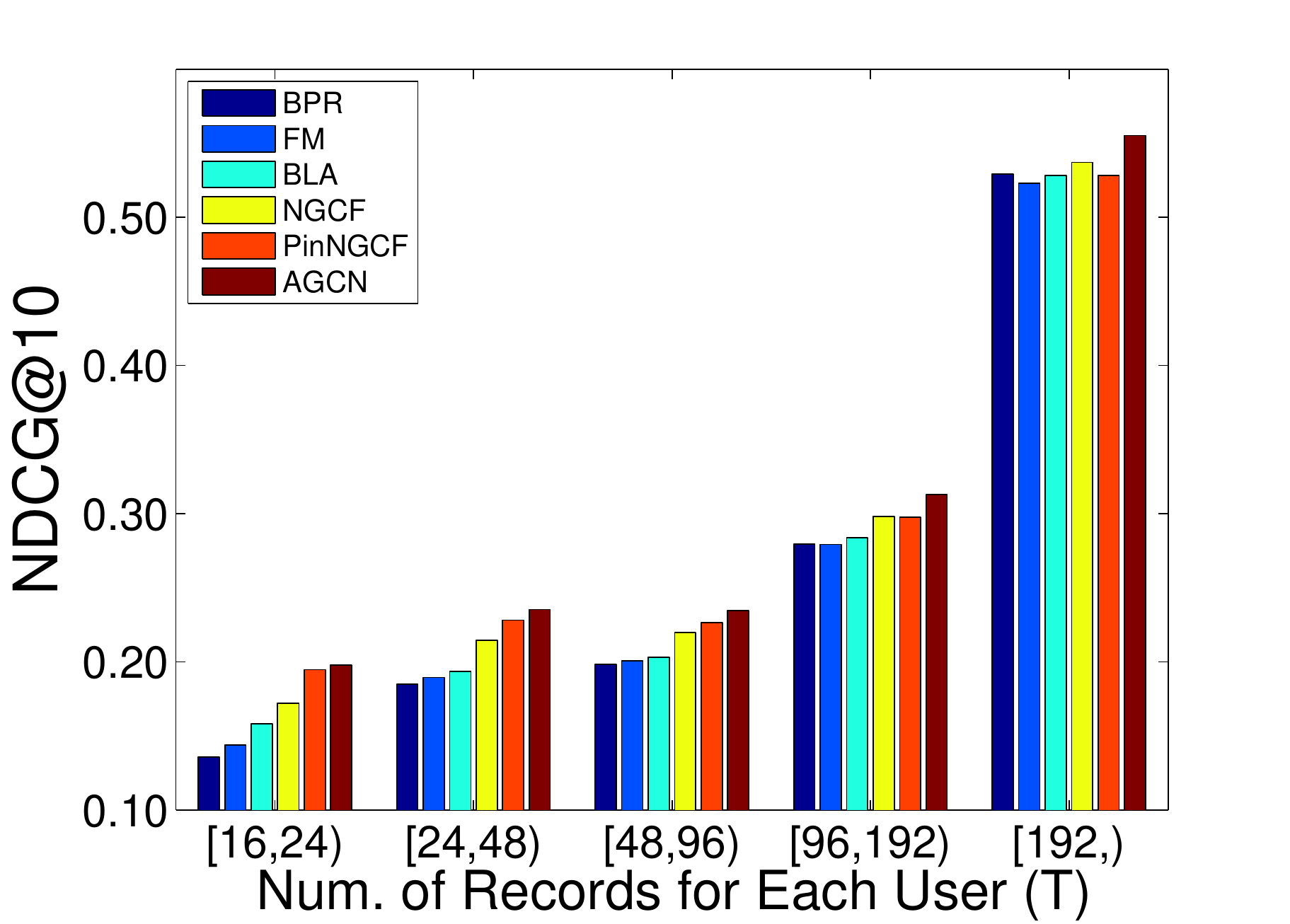}}
  \end{center}
  \vspace{-0.2cm}
  \caption{Item recommendation performance under different user group.} \label{fig:sparsity_results}
  \vspace{-0.2cm}
\end{figure*}

\subsubsection{Parameter Settings}
For all latent embedding based models, we initialize the embedding matrices with a Gaussian distribution with the mean value of 0 and the standard variance of 0.01.  For these gradient descent-based methods, we use Adam as the optimizing method with an initial learning rate of 0.001 and a batch size of 1024 during model learning. We stop the model learning process when the performance decreases in the validation data. In our proposed \shortname~model, we set embedding size $D$ in [16,32,64,128] and find $D=32$ reaches the best performance. We set the regularization parameter $\lambda$ in [0.1,0.01,0.001,0.0001], and find $\lambda=0.01$ reaches the best performance. Similar to many graph-based recommendation models~\cite{ICLR2017Semigcn,KDD2018PinSage}, we set the depth parameter $K$ in [0,1,2,3,4], and would analyze its impact in the experiments.
Since we have the rating based loss, at each iteration of the training process, for each observed user-item interaction, we randomly select one unobserved item as candidate negative sample to compose a triple data. Therefore, the candidate negative samples change in each iteration and give weak signal for model learning. There are several other parameters in the baselines, we tune all these parameters to ensure the best performance of the baselines for fair comparison.

\subsection{Performance on Item Recommendation}
\subsubsection{Baselines and Evaluation Metrics}
We compare our \shortname~model with following state-of-the-art baselines for item recommendation:
\begin{itemize}
    \item \textbf{BPR}~\cite{UAI2009BPR}: It is a competing latent factor model for implicit feedback based recommendation. It designed a ranking based loss function that assumes users prefer items they observed compared to unobserved ones.
	\item \textbf{FM}\cite{ICDM2010FM}: This model is a unified latent factor based model that leverages the user and item attributes. Since FM needs the complete user~(item) attributes, we complete the missing attribute values with the best attribute inference model in practice.
	\item \textbf{NGCF}~\cite{SIGIR2019NGCF}: It is a state-of-the-art graph based collaborative filtering model.  NGCF iteratively learns user and item representations from aggregating neighbors' embeddings in the previous layers.
	\item \textbf{PinNGCF}: We call this model as PinNGCF as it combines state-of-the-art content based graph model of PinSage~\cite{KDD2018PinSage} and the collaborative filtering based graph model NGCF~\cite{SIGIR2019NGCF}. Therefore, PinNGCF is a designed hybrid graph based recommendation model, in which each user~(item)'s input layer is a concatenation of the free embedding vector and the node's attribute vector.
	\item \textbf{BLA}~\cite{WWW2017BLA}: It is a joint model that combines link prediction and attribute inference on social graph. It assigns different weights to edges on the graph, then uses graph construction to make item recommendation.
\end{itemize}
\vspace{-0.2cm}

For item recommendation task, we focus on recommending top-N items for each user. We adopt two widely used ranking based metrics: Hit Ratio~(HR)~\cite{SIGIR2019DiffNet} and Normalized Discounted Cumulative Gain~(NDCG)~\cite{SIGIR2019DiffNet}. Specifically, HR measures the number of successfully predicted items in the top-N ranking list that the user likes in the test data. NDCG considers the hit positions of the items and will give a higher score if the hit items are in the top positions.
In practice, we select all unrated items as negative samples for each user, and combine them with the positive items the user likes in the ranking process.

\begin{table*}
	\begin{small}
		\centering
		\setlength{\belowcaptionskip}{5pt}
		\caption{Performance comparisons for attribute inference.}\label{tab:attribute inference}
		\begin{tabular}{|c|c|c|c|c|c|c|c|}
			\hline
			\multirow{2}{*}{Models} &
			\multicolumn{3}{|c|}{Amazon-Video Games}& \multicolumn{3}{|c|}{Movielens-1M} &\multicolumn{1}{|c|}{Movielens-20M} \\
			\cline{2-8}
			&Price(ACC)&Platform(MAP)&Theme(MAP)&Gender(ACC)&Age(ACC)&Occupation(ACC)&Genres(MAP)\\
			\hline
			LP&0.1987&0.4545&0.5772&0.7168&0.3463&0.1126&0.5654 \\
			\hline
			GR&0.1488&0.4956&0.6431&0.7211&0.3354&0.1127&0.5795 \\
			\hline
			Semi-GCN&0.1535&0.4968&0.6504&0.7164&0.3473&0.1267&0.5587\\
			\hline
			BLA&0.1637&0.6064&0.6537&0.7206&0.3561&0.1245&0.5960\\
			\hline
			\textit{\textbf{AGCN}}&\textbf{0.2083}&\textbf{0.7762}&\textbf{0.7294}&\textbf{0.7574}&\textbf{0.3953}&\textbf{0.1477}&\textbf{0.6443} \\
			\hline
		\end{tabular}
	\end{small}
\end{table*}

\subsubsection{Overall Performance}
Table~\ref{tab:hr@n_topk} and Table~\ref{tab:ndcg@n_topk} report the overall item recommendation results  with different top-N values on HR@N and NDCG@N metrics, respectively. We observe that all models outperform BPR since they either leverage additional data or use more advanced modeling techniques. Specifically, NGCF is the sole remaining baseline that only takes the user-item interaction behavior as input, and shows huge improvement compared to BPR.
Though most attribute values are incomplete, by filling these missing values with pre-trained models, the attribute enhanced models have better performance compared to their counterparts that do not leverage the attribute data.
In other words, FM improves over BPR, and PinNGCF further improves over NGCF. When comparing the two attribute enhanced recommendation models, the graph based model PinNGCF still outperforms FM. BLA is the only baseline that jointly predicts the two tasks. As BLA is based on classical attribute inference and recommendation model, BLA has better results than BPR and FM with mutual reinforcement learning of the two tasks. However, BLA does not perform as well as these GCN based baselines. As such, we could also empirically conclude the superiority of applying GCNs for recommendation, as GCNs could inject higher-order graph structure for better user~(item) embedding learning. This is also the reason why we use GCN as the base model for our proposed model.

When comparing our proposed \shortname~model with the baselines, we empirically find \shortname~ improves over all baselines on three datasets with different evaluation metrics. The detailed improvement rate varies accross different datasets, but the overall trend is same.
E.g., \shortname~ improves about 8\% and 7\% compared to the best baseline~(i.e., PinNGCF) of HR@10 and NDCG@10 on  Amazon-Video Games dataset. Since PinNGCF could be seen as a simplified version of AGCN without the attribute update and joint modeling step, the results suggest the superiority of AGCN with attribute update and joint modeling.

\subsubsection{Sparsity Analysis}
In this part, we show the item recommendation performance of various models under different data sparsity. We split all users into 5 groups according to the training records of this user, and test the NDCG@10 performance of different user groups.  The results are shown in Figure\ref{fig:sparsity_results}. As shown in this figure, the x-axis shows the user group information and the y-axis denotes the performance. E.g, the user group information with $[8,16)$ in the horizontal axis means each user in this group
has at least 8 rating records and less than 16 rating records. Please note that, the detailed
group rating information varies across different datasets as the three datasets have different sizes, we bin users into different groups to ensure each group has similar users. As observed from this figure, the performance of all models increases with more user records. Specifically, when the user group is very sparse, graph based models~(i.e., NGCF, PinNGCF and AGCF) show larger improvements compared to the remaining models. On Amazon dataset, FM, NGCF and PinNGCF improve over BPR by about 5.0\%, 21.1\%, 35.5\% and 42.5\%  on the first group of users, separately. As user records increase, the improvement rate decreases, but the trend of the superiority of different models is the same.

\begin{table}
	\begin{small}
		\centering
		\setlength{\belowcaptionskip}{5pt}
		\caption{Performance comparisons of different propagation depth $K$ on three datasets.}\label{tab:propagation layers}
        \scalebox{0.86}{
		\begin{tabular}{|c|c|c|c|c|c|c|c|}
			\hline
			\multirow{2}{*}{Depth} &
			\multicolumn{2}{|c|}{Amazon-Video Games}& \multicolumn{2}{|c|}{Movielens-1M} &\multicolumn{2}{|c|}{Movielens-20M} \\
			\cline{2-7}
			&HR@10&NDCG@10&HR@10&NDCG@10&HR@10&NDCG@10\\
			\hline
			K=0&0.0636&0.0351&0.2119&0.1946&0.2667&0.2514\\
			\hline
			K=1&0.0741&0.0394&0.2230&0.2055&0.2876&0.2782\\
			\hline
			K=2&0.0832&0.0453&\textbf{0.2261}&\textbf{0.2065}&\textbf{0.2992}&\textbf{0.2903}\\
			\hline
			K=3&\textbf{0.0861}&\textbf{0.0461}&0.2152&0.1921&0.2965&0.2853\\
			\hline
    		K=4&0.0846&0.0456&0.2029&0.1738&0.2725&0.2559\\
			\hline
		\end{tabular}}
	\end{small}
	\vspace{-6mm}
\end{table}

\subsection{Performance on Attribute Inference}

\subsubsection{Baselines and Evaluation Metrics}
We compare AGCN with the following state-of-the-art attribute inference baselines:

\begin{itemize}
    \item \textbf{LP}~\cite{zhu2002LP}: Label Propagation(LP) is a classical algorithm for inferring node attributes based on graph structure by propagating each node's attribute to its neighbors.
    This model does not model the correlation of different attributes.
    \item \textbf{GR}~\cite{belkin2006manifold}: Graph Regularization (GR) extends classical supervised learning algorithm  with an additional regularization term, in order to enforce connected nodes to share similar attribute values. In practice, the supervised model fill the
    missing attribute input with an average of its neighbors, and adopt a simple linear model as the prediction function.
    \item \textbf{Semi-GCN}~\cite{ICLR2017Semigcn}: It is a state-of-the-art graph convolutional network for semi-supervised classification. Since there are missing feature values in the  input of Semi-GCN, we complete these missing values with the average of neighbors' values.
    \item \textbf{BLA}~\cite{WWW2017BLA}: It is a joint model which combines user links and user attribute inference on social graph with classical graph learning models. It assigns different weights to edges on the graph, then uses label propagation to infer attributes.
\end{itemize}

For attribute inference task, we use \emph{ACC}uracy~(ACC)~\cite{deng2009imagenet} to evaluate the single-label attribute inference performance and use \emph{M}ean \emph{A}verage \emph{P}recision~(MAP)~\cite{moghaddam1995probabilistic} to evaluate the multi-label attribute inference performance.
Specifically, ACC describes the proportion of correctly predicted samples to total samples. MAP describes the average precision of the inferred attributes, and it is widely used in multi-label classification tasks~\cite{IJCV2010pascal,deng2009imagenet,mcauley2015image}. For both metrics, the larger value means the better performance.
We have to note that due to data limit, we could not perform both user attribute inference and item attribute inference on each dataset . Our proposed \shortname~ is flexible to predict either of them according to the available data.

\begin{figure*}
\vspace{-0.2cm}
  \begin{center}
  \vspace{-0.2cm}
      \subfigure[Amazon-Video Games \textit{Price}]{\includegraphics[width=45mm]{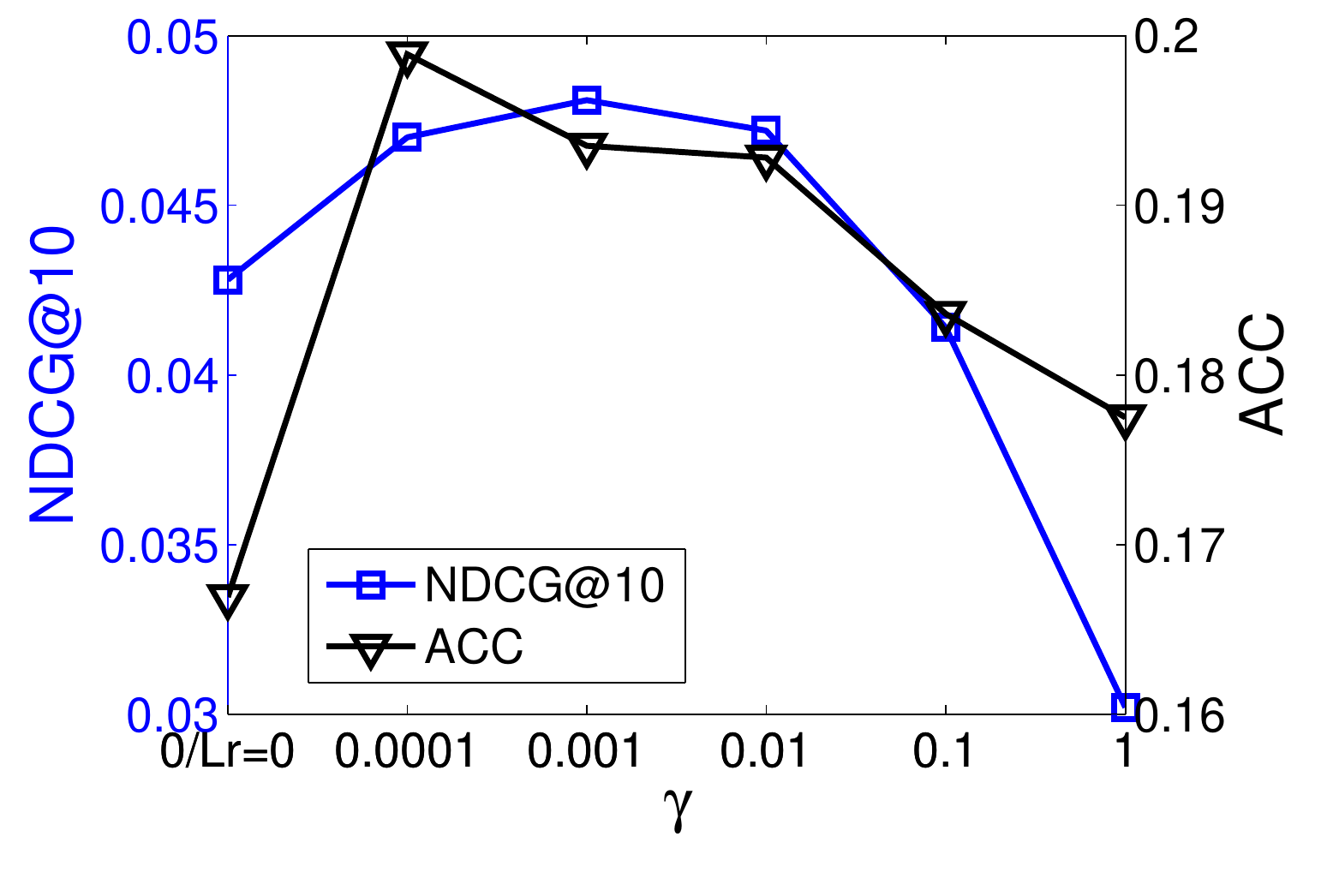}}
      \subfigure[Amazon-Video Games \textit{Platform}]{\includegraphics[width=45mm]{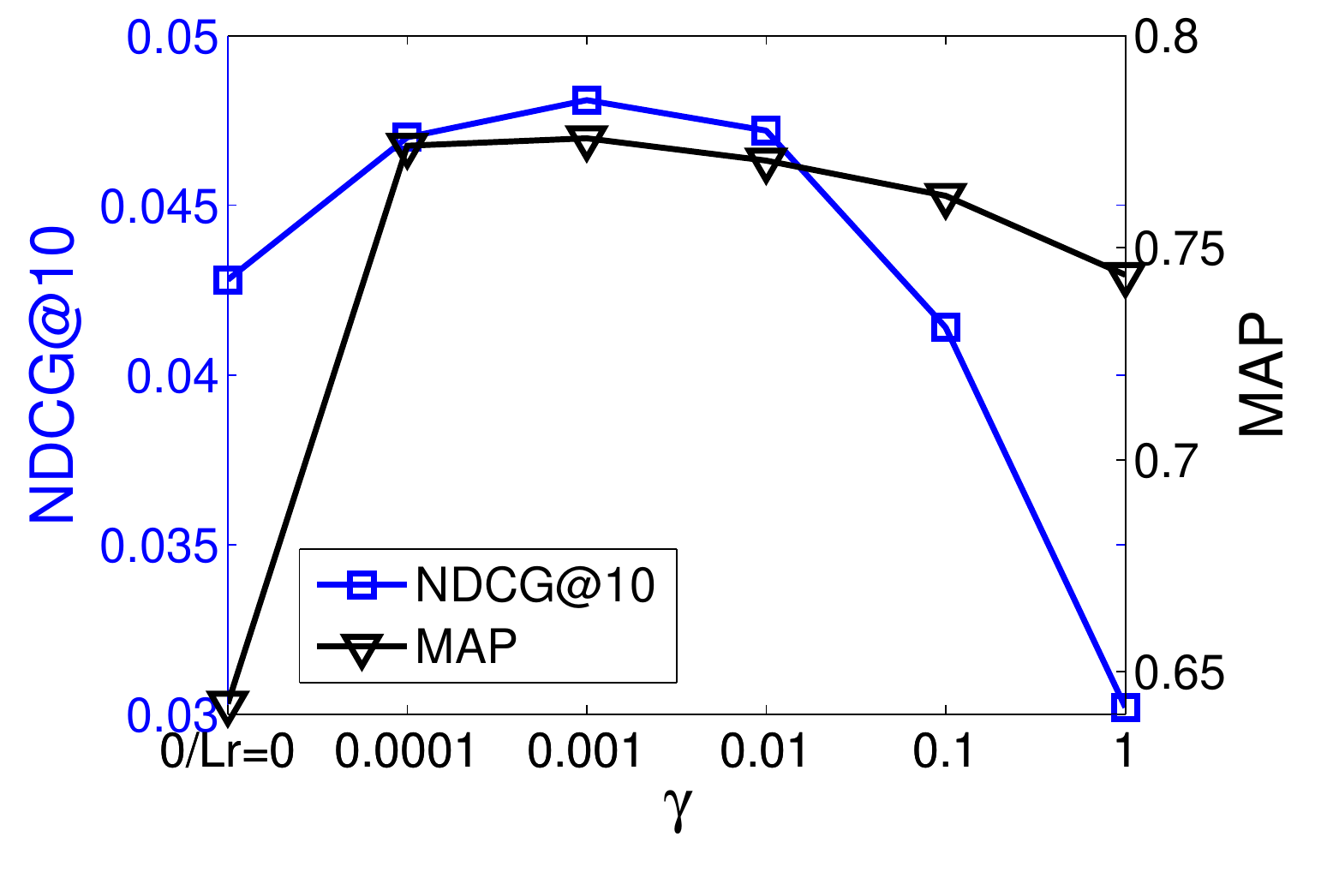}}
      \subfigure[Amazon-Video Games \textit{Theme}]{\includegraphics[width=45mm]{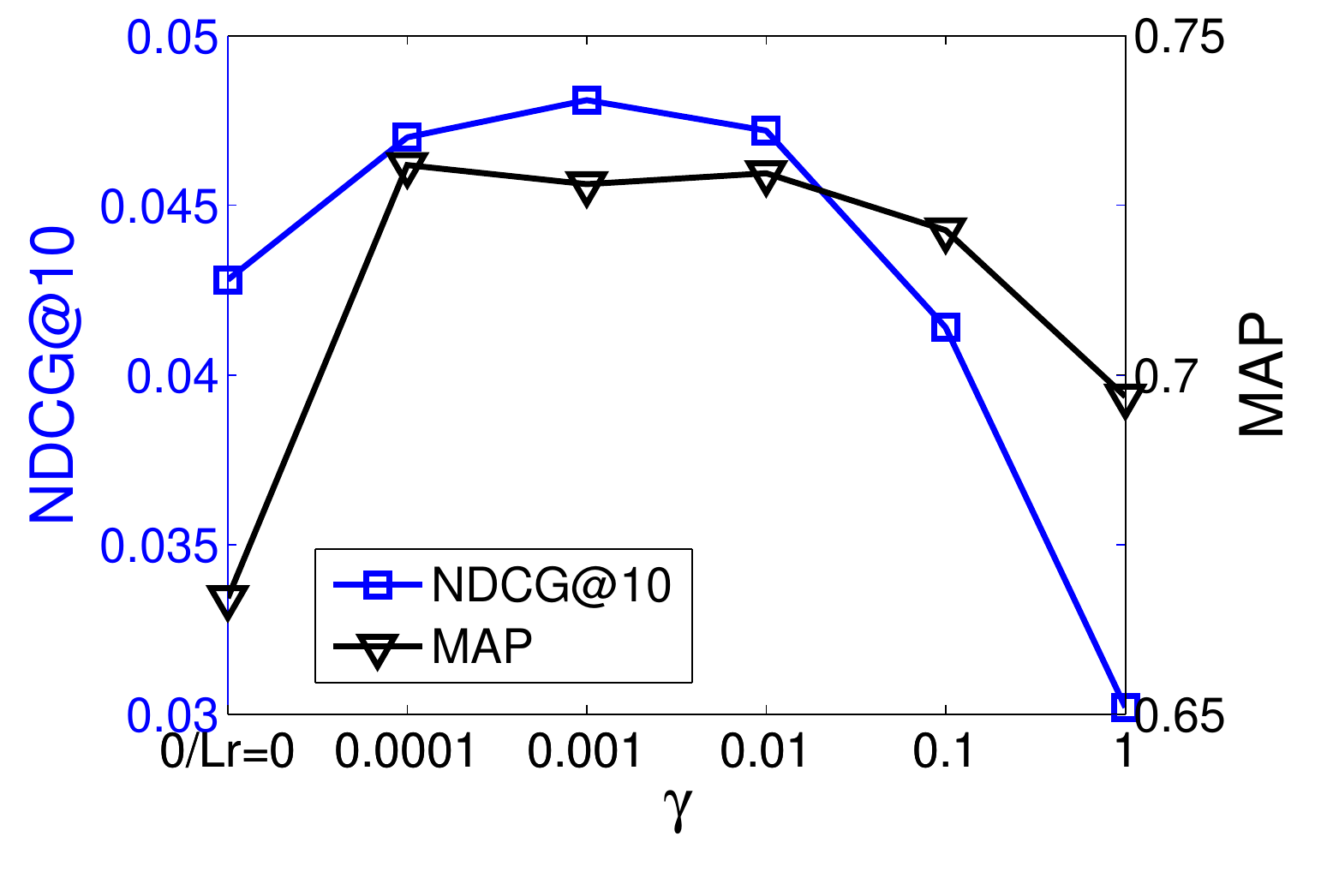}}
  \end{center}
  \vspace{-0.2cm}
  \vspace{-0.2cm}
\end{figure*}

\begin{figure*}
\vspace{-0.2cm}
  \begin{center}
  \vspace{-0.2cm}
      \subfigure[Movielens-1M \textit{Gender}]{\includegraphics[width=42mm]{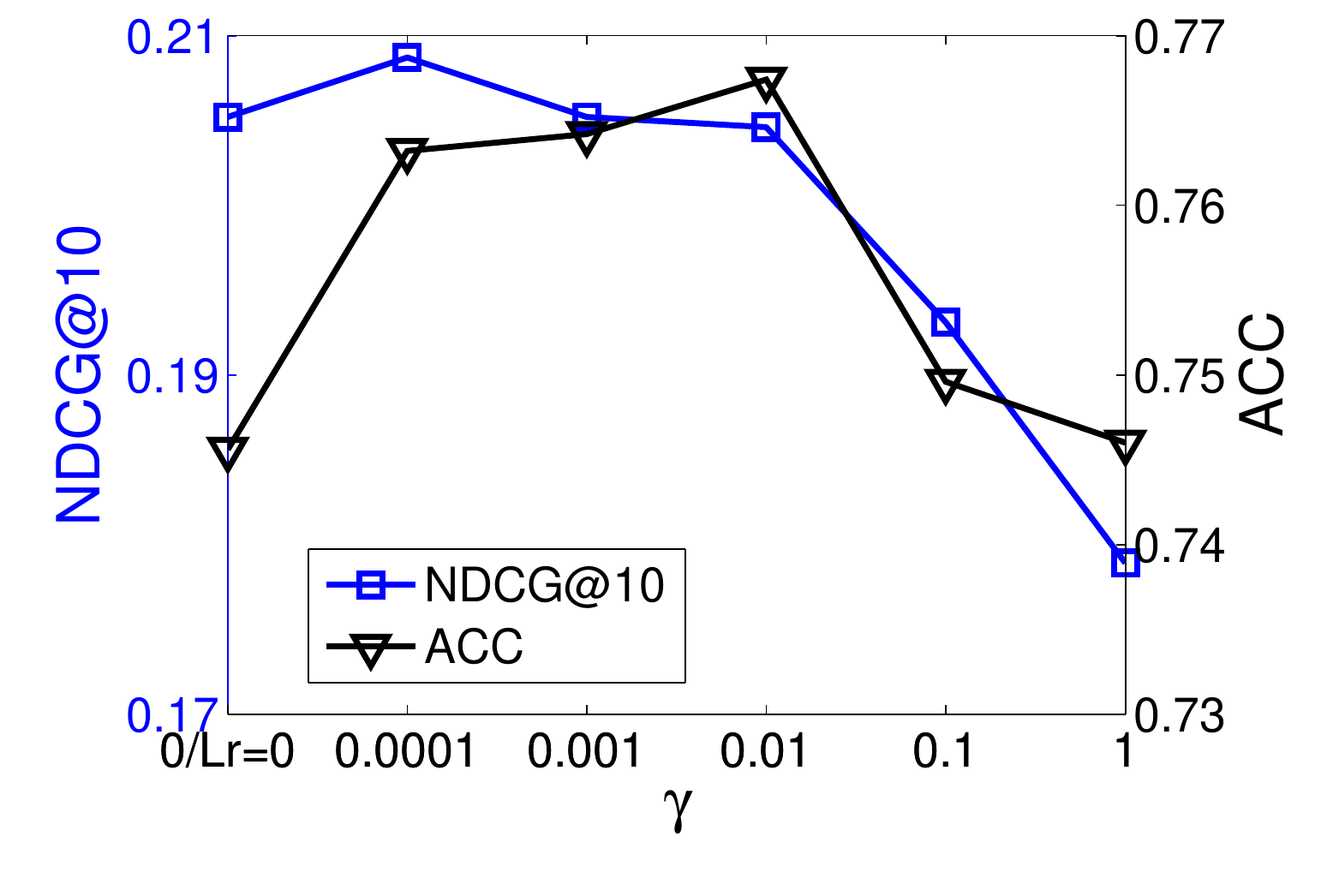}}
      \subfigure[Movielens-1M \textit{Age}]{\includegraphics[width=42mm]{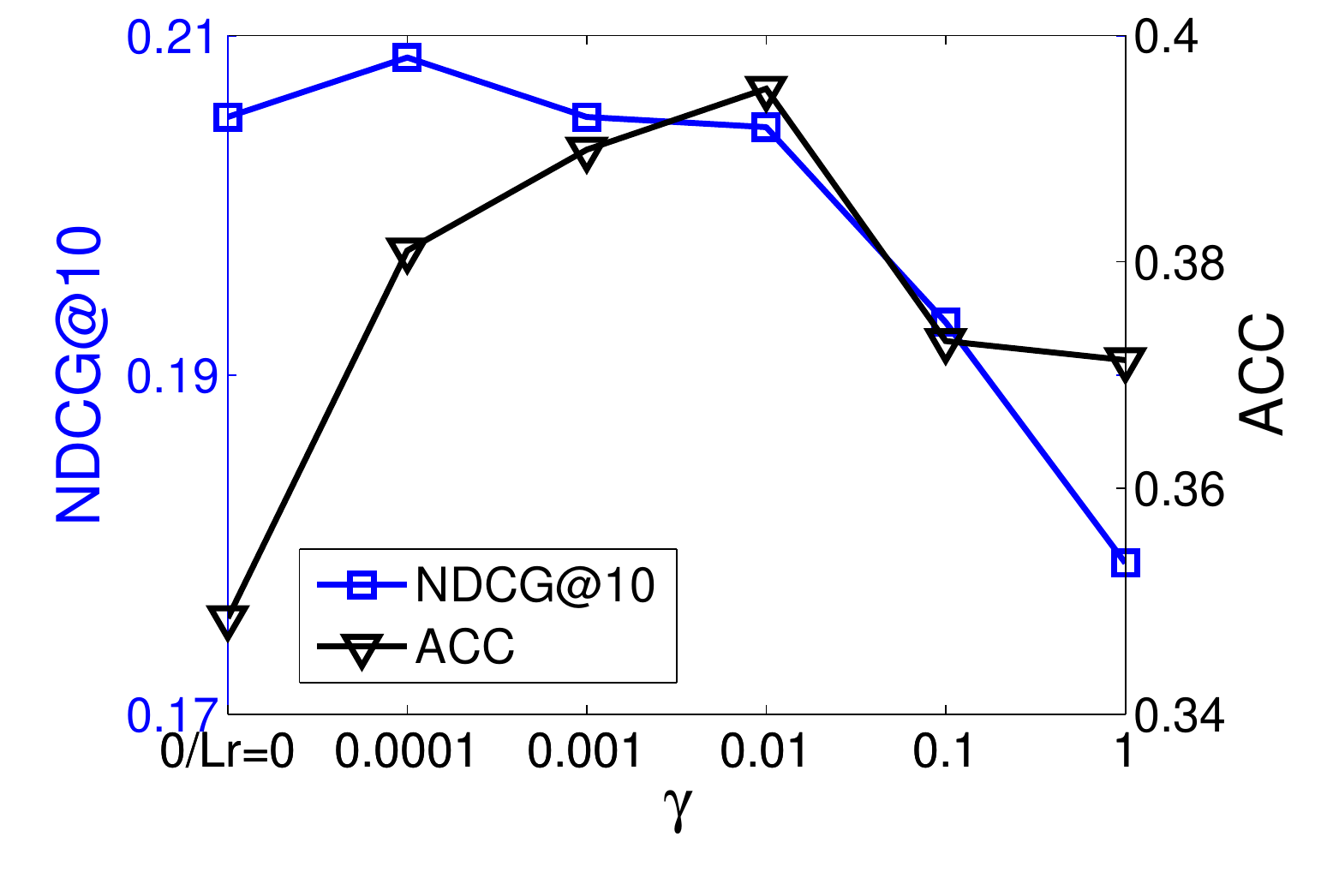}}
      \subfigure[Movielens-1M \textit{Occupation}]{\includegraphics[width=42mm]{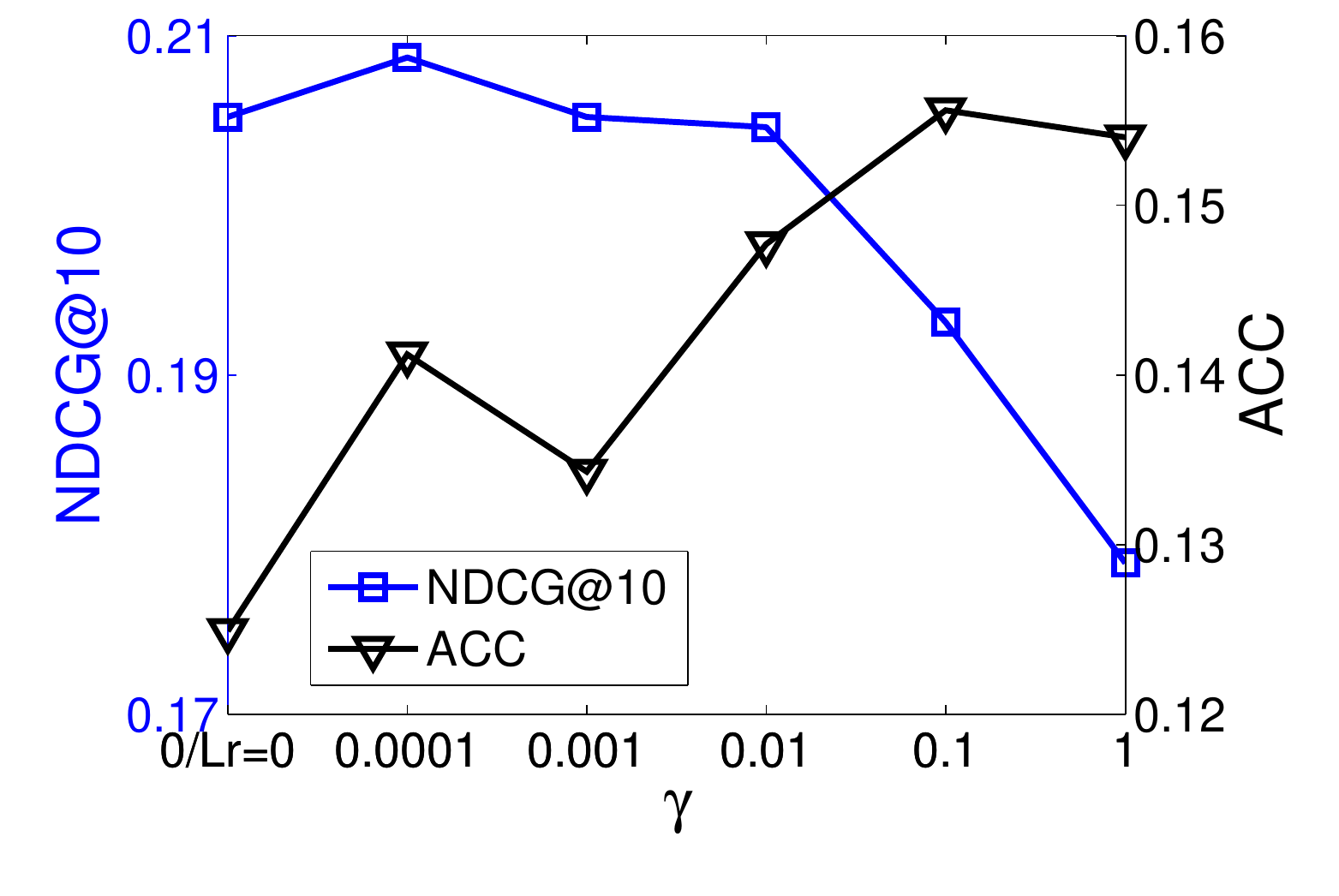}}
      \subfigure[Movielens-20M \textit{Genres}]{\includegraphics[width=42mm]{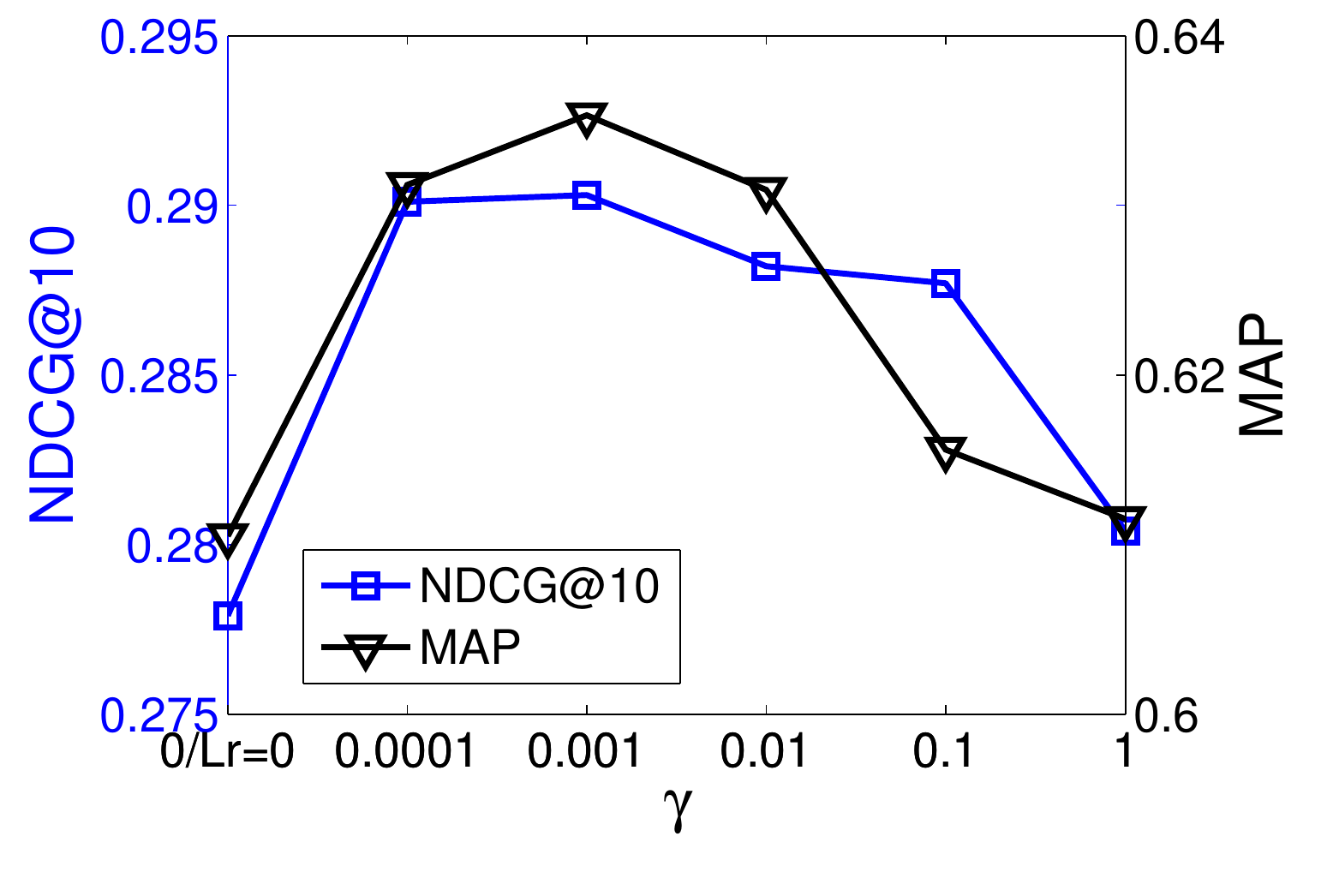}}
  \end{center}
  \vspace{-0.4cm}
  \caption{\small{NGCD@10 and different attribute inference performance under different $\gamma$ on three datasets}}
  \label{fig:analysis-gama}
  \vspace{-0.3cm}
\end{figure*}

\subsubsection{Overall Performance}
Table~\ref{tab:attribute inference} shows the performance of our model comparing to the baselines.
Our proposed model shows the best performance for each attribute inference on different dataset. For example,  our model improves over strongest baseline with 4.8\%, 28.0\% and 11.6\% on price, platform and theme attributes of Amazon-Video Games, separately. On Movielens-1M, our model improves over the strongest baseline with 5.1\%, 13.8\% and 16.4\% on gender, age and occupation attributes. On Movielens-20M, our model also improves over strongest baseline with 6.6\% on genres attribute.

When comparing the baselines  of LP, GR and Semi-GCN that are designed for attribute inference, we find GR and Semi-GCN show better performance than LP in most situations, as these models leverage the related remaining attributes in the modeling process. There is an exception of the price attribute prediction of Amazon-Video Games dataset, with LP is the strongest baseline. We guess a possible reason is that, price is not correlated to other attributes such as platform and theme, and introducing other attributes would add noisy information for the remaining baselines. GR and Semi-GCN show similar performance as both models fill the missing attribute values with precomputed data. Generally, BLA is the best baseline by jointly modeling the two tasks. Our proposed model could further improve BLA as AGCN better models the complex graph structure for to enhance both tasks.

\subsection{Detailed Model Analysis}

\subsubsection{The Propagation Layer Depth $K$.} Table~\ref{tab:propagation layers} summarizes the experimental results of AGCN with different propagation depth $K$. We set depth $K$ in the range of $\{0,1,2,3,4\}$, and report the values of HR@10 and NDCG@10 with various depth $K$. Specifically, when $K=0$, AGCN does not incorporate any graph structure for model learning. With the increase of $K$, the up to K-th higher order graph structure is leveraged for node representation. As shown in this table, the performance improves quickly on all datasets, as the first-order connected neighbors' information is crucial for alleviating the data sparsity issue. As $K$ continues to increase, the trend is that the performance still increases at first, but would drop after a value. Specifically, our model achieves the best performance with $K=3$ on Amazon-Video Games, $K=2$ and Movielens-1M and Movielens-20M. We guess that, Amazon-Video Games has very sparse rating records with only 0.028\% density, so more neighbors aggregation is better for node embedding learning. However, for Movielens-1M and Movielens-20M, too many 3-order neighbors would lead over smoothing on the graph, so the performance decreases compare to $K=2$. When $K$ continues increase to 4, \shortname  is already overfitting,  and the performance decreases on all datasets. The reason is that, the graph convolutions can be regraded as feature smoothing in the graph. When introducing more layers, the representation ability
is over smoothed~\cite{AAAI2018deeper}. In fact, this phenomenon has also been observed in other GCN based recommendation tasks~\cite{SIGIR2019DiffNet, SIGIR2019NGCF, KDD2018PinSage}.

\subsubsection{Task Balance Weight $\gamma$.} As shown in Eq.\eqref{eq:loss_all}, parameter $\gamma$ controls the relative weight of the two losses in \shortname, i.e., the rating loss function and the attribute inference loss.
The larger the  $\gamma$, the more we rely on the attribute inference loss for joint task prediction. Specifically, when $\gamma=0$, the attribute inference loss disappears and we only rely on the preference prediction loss for prediction. In this part, we would like to show how the task balance parameter $\gamma$ influences the performance of the two tasks. Specifically, as we aim at both item recommendation and attribute inference, $\gamma=0$ denotes we only use the rating based loss for item recommendation, and $\mathcal{L}_r=0$ denotes we only consider the attribute inference loss for attribute inference. Therefore, the correlations of these two tasks are not well captured, and the two tasks are modeled independently under such setting. The performance of the two tasks with different balance parameteres  is shown in Figure~\ref{fig:analysis-gama}, with each attribute inference result is shown in the right y-axis, and the item recommendation performance is displayed in the left y-axis. As can be observed from each subfigure, the left most part of x-axis shows the performance without any task correlation modeling~( $\gamma=0$ for item recommendation  and $\mathcal{L}_r=0$ for attribute inference). As $\gamma$ increases from 0 to 0.001, the performance of the two tasks are much better than the results at the leftmost part of each subfigure, showing the soundness of modeling these two tasks in a unified framework. As $\gamma$ continues to increase, the performance of the two tasks increases at first, and drops when $\gamma=0.1$ or $\gamma=1$ for most subfigures. Different tasks achieve the best performance under different $\gamma$. Besides, different kind of attributes also achieve best performance with different $\gamma$. In practice, we choose the best $\gamma$ parameter for each task.

\section{CONCLUSION}
In this paper, we proposed a \shortname~ model for joint item recommendation and attribute inference in an attributed user-item bipartite graph with missing attribute values. To tackle the missing attribute problem, AGCN was designed to iteratively  performing two steps: graph embedding learning with previous learned attribute values, and attribute update procedure to update the input of graph embedding learning. Therefore, AGCN could adaptively adjusted the graph learning process by incorporating the given attributes and the estimated attributes, in order to provide weak supervised signals to facilitate both tasks. Experimental results on three real-world datasets clearly showed the effectiveness of our proposed model.

\vspace{-0.1cm}
\section*{Acknowledgements}

This work was supported in part by National Key Research and Development Program of China(Grant No.2017YFB0803301), the National Natural Science Foundation of China(Grant No.61725203, 61972125, U19A2079, 61722204, 61932009 and 61732008).

\bibliographystyle{ACM-Reference-Format}
\bibliography{attribute}

\appendix
\end{document}